\documentclass[twoside,12pt]{article}

\usepackage{amssymb,endnotes}

\newcounter{muni}
\newenvironment{remunerate}{\begin{list}{{\rm \arabic{muni}.}}
{\usecounter{muni}\setlength{\leftmargin}{0pt}
\setlength{\itemindent}{38pt}}}{\end{list}}

\newcommand{\nc}{\newcommand}                  \nc{\nf}{\infty}    
\nc{\dst}{\displaystyle}                       \nc{\nnb}{\nonumber} 
\nc{\beq}{\begin{equation}}                    \nc{\eeq}{\end{equation}} 
\nc{\beqa}{\begin{eqnarray}}                   \nc{\eeqa}{\end{eqnarray}}
\nc{\brm}{\begin{remunerate}}                  \nc{\erm}{\end{remunerate}}
\nc{\barr}{\begin{array}}                      \nc{\earr}{\end{array}}
             \nc{\mc}{\mathcal}

\nc{\bs}{\backslash}        \nc{\nl}{\newline}      \nc{\mb}{\mathbb}
\nc{\qq}{\quad\quad}        \nc{\ol}{\overline}     \nc{\pt}{\partial}
\nc{\dg}{\dagger}

\nc{\alf}{\alpha}    \nc{\be}{\beta}        \nc{\ga}{\gamma}   \nc{\si}{\sigma} 
\nc{\de}{\delta}     \nc{\eps}{\epsilon}    \nc{\vtht}{\vartheta}
\nc{\om}{\omega}     \nc{\vp}{\varphi}    \nc{\vsi}{\varsigma}
\nc{\vrho}{\varrho}  \nc{\tht}{\theta}      \nc{\la}{\lambda}

\nc{\Om}{\Omega}     \nc{\Ga}{\Gamma}      \nc{\De}{\Delta}

\nc{\Log}{{\rm Log }}           \nc{\tg}{{\rm tg }}
\nc{\sh}{{\rm sh }}             \nc{\ch}{{\rm ch }}

\nc{\tr}{{\rm tr }}

\nc{\LIM}{\mathop{\smash{\rm LIM}}}

\title{ \bf Quantum structure of T-dualized models with symmetry breaking}
\author{Pierre-Yves CASTEILL\thanks
{\noindent Laboratoire de Physique Th\'eorique et des Hautes Energies,
 Unit\'e associ\'ee au CNRS UMR 7859,~Universit\'e Paris 7,
 2 Place Jussieu, 75251 Paris Cedex 05.}  
\and Galliano VALENT $\!\!^{\ast}$}
\date{ }
\begin{document}
\maketitle

\vspace{-7cm}
\begin{flushright}
LPTHE 00-24 \\
hep-th/0006186 \\
June 2000
\end{flushright}
\vskip 1.0truecm

\vspace{6cm}

\begin{abstract}
We study the principal $\si$-models defined on any group manifold 
$\,G_L\times G_R/G_D$ with breaking of $\,G_R,$ and their T-dual 
transforms. For arbitrary breaking we can express the torsion and 
Ricci tensor of the dual model in terms of the frame geometry of the initial 
principal model. Using these results we give necessary and sufficient 
conditions for the dual model to be torsionless and prove that the one-loop 
renormalizability of a given principal model is inherited by its dual partner, 
who shares the same $\,\be\,$ functions. These results are shown to hold also 
if the principal model is endowed with torsion. 
As an application we compute the $\,\be$ functions for the full Bianchi 
family and show that for some choices of the breaking parameters the dilaton 
anomaly is absent : for these choices the dual torsion vanishes. For the 
dualized Bianchi V model (which is torsionless for any breaking), we take 
advantage of its simpler structure, to study its two-loops renormalizability.
\end{abstract}

\newpage

\section{Introduction}
The subject of classical versus quantum equivalence of T-dualized $\si$-models 
has been strongly studied in recent years, and extensive reviews covering 
abelian, non-abelian dualities and their applications to string theory and 
statistical physics are available \cite{aab1},\cite{aal3},\cite{gpr}. More 
recent developements on the geometrical aspects of duality can be found 
in \cite{Al1}. After the proof that T-duality is indeed a canonical 
transformation \cite{Lo}, \cite{Sf} relating two classically equivalent 
theories, the most interesting 
problem is to study this equivalence at the quantum level. This was done 
mostly for dualizations of Lie groups, with emphasis put on $\,SU(2).$ For 
this model the one-loop equivalence was established in \cite{fj}, \cite{ft}.

The way towards the general case was cleaned up with the derivation of the 
classical structure of the non-abelian dual for any group \cite{ft}, 
\cite{aal1}, \cite{aab1}, \cite{gr}. However the analysis of Bianchi V in 
\cite{grv} revealed that for some renormalizable dual theories the 
divergences could not be absorbed by a re-definition of the dilaton field! It 
was further realized that this phenomenon occurs for non semi-simple Lie 
groups with traceful structure constants ($f^s_{si}=0$), and that it can be 
interpreted as a mixed gravitational-gauge anomaly \cite{aal1}.

A further decisive progress was made by  Tyurin \cite{Ty}, who generalized 
the one-loop equivalence to an arbitrary Lie group and derived the general 
structure of the dilaton anomaly. However, as pointed out in \cite{bbfm}, his analysis 
considers only models with explicit invariance under the left group action (whose existence is crucial for the dualization process) leaving aside the right action 
and the possible symmetry breaking schemes for it. The one-loop equivalence problem 
in this more general setting has been examined recently \cite{bbfm},\cite{hk1} 
for the group manifold $SU(2)_L\times SU(2)_R/SU(2)_D$, where $SU(2)_R$ is 
broken down to a $U(1).$ The renormalizability and dilatonic properties do 
survive despite the lowering of the right isometries. It is the purpose of 
the present article to analyze the geometry of the dualized model for a large 
class of models built on  
$G_L\times G_R/G_D,$ with arbitrary breaking of $G_R.$ While in \cite{Ty} 
supersymmetry considerations \`a la Busher \cite{Bu1},\cite{Bu2} were convenient  
to derive the dualized geometry, we will show that a direct 
computation in local coordinates is fairly efficient to extract 
the Ricci tensor in the presence of symmetry breaking.

The content of this article is the following : after setting the notations , in 
section 2 we study the geometry of the group manifold 
$(G_L\times G_R)/G_D.$ This is most conveniently done using frames and, despite 
symmetry breaking, one obtains a manageable form for the Ricci tensor. In section 3 
the dualized theory is examined and its torsion and Ricci tensor are  
computed, exhibiting their dependence with respect to the geometrical 
quantities of the principal 
model. The possibility of torsionless dualized models is discussed. 
In section 4 we use the previous results to show that the one-loop renormalizability of
 the principal model is inherited by its T-dual. In section 5 we generalize the 
previous analyses to deal with a principal model endowed with torsion. 
In section 6 we examine the models 
in the Bianchi class, compute their beta functions, and for the non semi-simple 
algebras discuss the dilaton anomaly. For some breaking choices this anomaly may 
vanish and in these cases the dual models are torsionless. Since any dualized 
Bianchi V model is torsionless, we study in section 7, for the simplest breaking, 
its two-loops renormalizability.

\section{Geometry of the broken principal models}
Since we have in view perturbative applications, our considerations will be 
of a local nature. Let us consider a Lie algebra 
$\,{\cal G}=\{X_i,\ i=1,\cdots,\nu\}\,$ with structure constants
$$[X_i,X_j]=f_{ij}^{~s}\,X_s.$$
Denoting by $\,z^i$ the local coordinates in a neighbourhood of the origin, we 
exponentiate to the group by $\,g=\exp(z\cdot T),$ and define
\beq\label{i2}
g^{-1}\pt_{\mu}g=J_{\mu}^i\,X_i.\eeq
For further use we introduce the adjoint representation by
\beq\label{repadj}
(T_i)_{j}^{~k}\equiv ({\rm ad}\,X_i)_j^{~k}=-f_{ij}^{~k},\eeq
which allows to write the Jacobi identity
\beq\label{j}
[T_i,T_j]=f_{ij}^{~s}T_s,\qq\quad i,j,s=1,\ldots\nu={\rm dim}\,({\cal G}).\eeq

Then the action of the  corresponding principal model can be written
\beq\label{i1}
S=\frac 12\int\,d^2x\,B_{ij}\,\eta^{\mu\nu}J_{\mu}^i\,J_{\nu}^j,\eeq
where the matrix $\,B$ is symmetric and invertible. For field theoretic 
applications one should add the restriction that $\,B\,$ is positive definite 
\cite{bbd}, while this does not seem to be necessary for stringy 
applications. This restriction implies, in the semi-simple case that its 
simple components have to be compact. Our analysis will not make use of this 
positivity hypothesis.

Taking the curl of the first relation in (\ref{i2}) gives the Bianchi identity
\beq\label{b}
M_{\mu\nu}^i(J)\equiv \pt_{\mu}J^i_{\nu}-\pt_{\nu}J^i_{\mu}
+f_{st}^{~i}J_{\mu}^s\,J_{\nu}^t=0\qq\Longleftrightarrow\qq 
\eps^{\mu\nu}M_{\mu\nu}^i(J)=0.\eeq

\subsection{Isometries}
Let us proceed to a discussion of the isometries of the action (\ref{i1}). The 
groups $G_L\times G_R$ and $G_D$ act on $g$ according to
\beq\label{g2}
g\quad\longrightarrow\quad g'=G_L\,g\,G_R^{-1},\qq\qq g\quad\longrightarrow\quad 
g'=G_D\,g\,G_D^{-1}.\eeq
As a consequence
$$g^{-1}\pt_{\mu}g\quad\longrightarrow\quad G_R\,g^{-1}\pt_{\mu}g\,G_R^{-1},$$
and specializing to infinitesimal transformations one gets
\beq\label{g3}
G_R\approx{\mb I}+\eps_R^iT_i,\qq\qq\Longrightarrow\qq\qq
\de J_{\mu}^k=f_{ij}^{~~k}\eps_R^iJ_{\mu}^j.\eeq
It follows that the action (\ref{i1}) is invariant under $G_L,$ while the 
matrix $B_{ij}$ will generally break $G_R$ down to some subgroup $H$ 
(possibly trivial). Denoting by $\{T_s,\ s=1,\ldots, h\}\,$ the generators of 
its Lie algebra $\,{\cal H},$ these should satisfy
\beq\label{g4}
(T_s)_i^{~k}B_{kj}+(T_s)_{j}^{~k}B_{ik}=0,\qq\quad\forall\  T_s\in {\cal H}.\eeq
Let us emphasis that the metric $\,B$ can be freely chosen (as far as it is 
symmetric and invertible!), but, if $\,{\cal G}$ is simple, 
the most symmetric choice is given by the bi-invariant metric
\beq\label{g5}
B_{ij}=\frac 1{\rho}g_{ij},\qq\qq g_{ij}={\rm Tr}\,(T_iT_j)=\tilde{\rho}\,
{\rm Tr}\,(t_it_j),\eeq
where $g_{ij}$ is the Killing metric and the $t_i$ the defining representation 
of the simple algebra under consideration. In the simple compact case we have

\vspace{5mm}
\centerline{
\begin{tabular}{|c|c|c|c|}
\hline
  & \quad $so(n)$ & \quad $su(n)$ & \quad $sp(n)$\\
\hline
$\tilde{\rho}$ & \quad $(n-2$) & \quad $2n$ &\quad $2(n+1)$\\
\hline
\end{tabular}}
\vspace{5mm}
\noindent and with the standard normalization of the generators 
$\,{\rm Tr}\,(t_it_j)=-2\de_{ij},$ we see that the choice 
$\rho=-2\tilde{\rho}$ gives $\,B_{ij}=\de_{ij}.$ In the simple non-compact 
case the same choice of $\,\rho\,$ gives $B_{ij}=\eta_{ij},$ which is diagonal, 
with $\,\eta_{ii}=+1$ for a compact generator $\,t_i\,$  and $\,\eta_{ii}=-1\,$
for a non-compact one.

The  bi-invariant metric has for isometry group the full $G_L\times G_R$ because 
$(T_s)_i^{~k}g_{kl}=-f_{sil}$ is fully skew-symmetric and therefore 
(\ref{g4}) is true for all the generators of $G_R.$

For a semi-simple $\,{\cal G}\,$ the situation is not very different, since 
it can be split into a direct sum of simple algebras
$${\cal G}={\cal S}_1\oplus\cdots\oplus{\cal S}_k,\qq\quad\qq
[{\cal S}_i,{\cal S}_j]=0\qq i\neq j.$$

\subsection{Geometry of frames}
In order to have a better insight into the geometry of the principal models 
with action (\ref{i1}), it is convenient to use a vielbein formalism, 
through the identification
$$B_{ij}\,\eta^{\mu\nu}\,J_{\mu}^i\,J_{\nu}^j
\qq\longleftrightarrow\qq B_{ij}e^ie^j,$$
and now the Bianchi identities appear as the Maurer-Cartan equations
\beq\label{MC}
de^i+\frac 12f_{st}^{~~i}e^s\wedge e^t=0.\eeq
We follow the notations of \cite{egh} and define the spin-connection  
$\om^i_{~j}$ by
$$de^i+\om^i_{~s}\wedge e^s=0,\qq\qq \om^i_{~j}=\om^i_{~j,s}e^s.$$
The frame indices are lowered or raised using the metric $B_{ij}$ and its 
inverse $B^{ij}=B^{-1}_{ij}.$ A straightforward computation gives
\beq\label{g6}
2\om_{ij,k}=f_{ij,k}+f_{ik,j}-f_{jk,i}\qq \qq
f_{ij,k}=f_{ij}^{~~s}B_{sk}.\eeq
For further use let us point out two consequences
\beq\label{g7}
\om^i_{~j,k}-\om^i_{~k,j}=-f_{jk}^{~~i},\qq\qq \om^s_{~i,s}=-f_{is}^{~~s}.\eeq
The curvature and the Ricci tensor are defined by
$$R^i_{~j}=d\om^i_{~j}+\om^i_{~s}\wedge\om^s_{~j}=
\frac 12 R^i_{~j,st}\,e^s\wedge e^t,\qq\quad ric_{ij}=R^s_{~i,sj}.$$
It follows that
\beq\label{g8}
R^i_{~j,st}=-\om^i_{~j,a}f_{st}^{~~a}-\om^i_{~a,t}\,\om^a_{~j,s}
+\om^i_{~a,s}\,\om^a_{~j,t}.\eeq
In the Ricci tensor the first two terms are gathered using (\ref{g7}) 
and give
\beq\label{g9}
ric_{ij}=-\om^s_{~i,t}\,\om^t_{~j,s}+\om^t_{~s,t}\,\om^s_{~i,j}.\eeq
The $i\,\leftrightarrow\,j$ symmetry of the first term is obvious while for 
the second it follows from
\beq\label{id1}
\om^t_{~s,t}(\om^s_{~i,j}-\om^s_{~j,i})=f_{st}^{~~t}f_{ij}^{~~s}=0,\eeq
where the last equality is obtained by taking the trace of the Jacobi identity 
(\ref{j}).

One can give the following explicit form of the Ricci tensor
\beq\label{ricci}
\barr{l}
ric_{ij}=\frac 12\,B_{st}\,(A^sB^{-1}A^t)_{ij}-
\frac 14\,B_{is}\,{\rm Tr}\,(B^{-1}A^sB^{-1}A^t)\,B_{tj}\\[3mm]
\hspace{3cm}-\frac 12\,{\rm Tr}\,(T_iT_j)+\frac 12\,{\rm Tr}\,(T_s)\,
\,\left(f^s_{~i,j}+f^s_{~j,i}\right),\qq f^s_{~i,j}=(B^{-1})_{st}f_{ti,j},
\earr\eeq
which exhibits that it is an homogeneous function of degree 0 in the 
breaking matrix $B.$ The scalar curvature $\,R=(B^{-1})_{ij}\,ric_{ij}\,$ is 
a constant, as it should for homogeneous spaces.

A drastic simplification takes place for the bi-invariant metric (\ref{g5}), for 
which we have
\beq\label{E}
ric_{ij}=-\frac{\rho}{4}\,B_{ij}.\eeq
The metric is therefore Einstein, and such a simple structure will have a 
counterpart in the dualized theory.

\subsection{Dualization}
For the reader's convenience we present a quick derivation 
\cite{ft},\cite{aab1} of the dualized model. The essence of the dualization 
process is to switch from 
the coordinates on the group, which parametrize $\,g,$ to new coordinates 
$\psi_i$ defined as the Lagrange 
multipliers of the Bianchi identities. Concretely this transformation is 
carried out starting from the action
$$S=\frac 14\int\,d^2x\left\{ B_{ij}\,\eta^{\mu\nu}J_{\mu}^i\,J_{\nu}^j
-\eps^{\mu\nu}\psi_iM_{\mu\nu}^i(J)\right\}.$$
Using light-cone coordinates, with the following conventions
$$x_{\pm}=\frac{x^0\pm x^1}{\sqrt{2}},\qq\eps_{01}=1,\qq
\eps^{\mu\si}\eps_{\si\nu}=\de^{\mu}_{\nu},\qq 
J_{\pm}=\frac{J_0\pm J_1}{\sqrt{2}},$$
one has
\beq\label{i3}
S=\frac 12\int\,d^2x\left\{(B+A\cdot\psi)_{ij}\,J^i_+J^j_-
-\psi_i(\pt_+J^i_-+\pt_-J^i_+)\right\}\eeq
with
\beq\label{i4}
(A^s)_{ij}=(T_i)_j^{~s}=-f_{ij}^{~s},\qq (A\cdot\psi)_{ij}=
(A^s)_{ij}\psi_s.\eeq
The field equations obtained from the variations with respect to the 
currents $J^i_{\pm}$ give
$$J^i_-=(B+A\cdot\psi)^{is}\,\pt_-\psi_s,\qq 
J^i_+=-\pt_+\psi_s\,(B+A\cdot\psi)^{si},\qq 
(B+A\cdot\psi)^{is}(B+A\cdot\psi)_{sj}=\de^i_j.$$
Using minkowskian coordinates on the worldsheet one has
$$J^{\mu i}=B^{ij}\,\eps^{\mu\nu}\left(\pt_{\nu}\psi_j
-(A\cdot\psi)_{jk}\,J_{\nu}^k\right),\qq\quad B^{is}B_{sk}=\de^i_k.$$
Using this relation, the action (\ref{i3}) can be written, up to total 
derivatives
\beq\label{d0}
S=\frac 12\int\,d^2x\,\pt_+\psi_i\,J^i_-=\frac 12\int\,d^2x\,\pt_+\psi_i
\,(B+A\cdot\psi)^{ij}\pt_-\psi_j.\eeq
Comparing this action with the one given in relation (4.16) of \cite{Ty} we 
see that in this reference only the unbroken case $\,B_{ij}=\de_{ij}\,$ 
has been considered.

Let us emphasize the following points :
\brm
\item Before dualization, all the field dependence on the coordinates chosen 
to parametrize $\,G\,$ must be hidden in expressions involving solely the 
currents $\,J_{\mu}^i.$ If this is not the case the dualization process 
is not possible.
\item The dualized action is completely defined by the breaking matrix $\,B$ 
and the field matrix $\,A\cdot\psi\ \in\ so(\nu).$ There are as many 
coordinates as generators in $\,{\cal G}.$
\item In the process of dualization the isometries corresponding to $\,G_L$ 
(which leave the $\,J_{\mu}^i$ invariant) are lost. This has for 
consequence that starting 
from an homogeneous metric, we are led to a non-homogeneous one.
\erm

\section{Geometry of the dualized theory}
In (\ref{d0}) we come back to standard 
notations and  change the coordinates $\psi_i$ to $\psi^i.$ Let us write 
the dual action
\beq\label{d1}
S=\frac 12\int\,d^2x\,G_{ij}\,\pt_+\psi^i\,\pt_-\psi^j,\qq 
G_{ij}=(B+A\cdot\psi)^{-1}_{ij}.\eeq
For further use we define the matrices
$$G^{\pm}=(B\pm A\cdot\psi)^{-1},\qq\quad G\equiv G^{+},
\qq\quad\Ga^{\pm}=B\pm A\cdot\psi,\qq (A\cdot\psi)_{ij}=-f_{ij}^{~s}\psi^s.$$
Writing the dual action (\ref{d1}) in minkowskian coordinates
\beq\label{d2}
S=\frac 12\int\,d^2x\left\{g_{ij}\,\eta^{\mu\nu}\pt_{\mu}\psi^i\pt_{\nu}\psi^j+
h_{ij}\,\eps^{\mu\nu}\pt_{\mu}\psi^i\pt_{\nu}\psi^j\right\},\eeq
gives for metric and torsion potential
$$g_{ij}=\frac 12(G_{ij}+G_{ji}),\qq\quad h_{ij}=
\frac 12(G_{ij}-G_{ji}),\qq\quad G_{ij}=g_{ij}+h_{ij}.$$
Using matrix notations we have
\beq\label{d3}
g=G^+\,B\,G^-=G^-\,B\,G^+,\qq\quad h=-g\,(A\cdot\psi)\,B^{-1}, \eeq
and for the inverse metric :
\beq\label{d4}
g^{-1}=\Ga^+B^{-1}\Ga^-=\Ga^-B^{-1}\Ga^+.\eeq
The determinant of the metric is
$$\det g=\frac{\det B}{(\det\Ga^{\pm})^2}=\det B\cdot(\det G^{\pm})^2.$$

\subsection{Connection}
We work with the standard conventions
\beq\label{torsion}
\Ga^i_{jk}=\ga^i_{jk}+T^i_{jk},\qq T^i_{jk}=g^{is}T_{sjk},\qq 
T_{ijk}=\frac 12(\pt_i h_{jk}+\pt_k h_{ij}+\pt_j h_{ki}),
\qq \pt_i\equiv\frac{\pt}{\pt\psi^i},\eeq
or using differential forms
$$H=\frac 1{2!}\,h_{ij}\,d\psi^i\wedge d\psi^j,\qq\qq 
T=\frac 1{3!}\,T_{ijk}\,d\psi^i\wedge d\psi^j\wedge d\psi^k=\frac 12\, dH.$$
The torsion potential is not uniquely defined since the following gauge 
transformation leaves invariant the torsion :
\beq\label{pdT}
H\to H+dA,\qq A=A_i\,d\psi^i,\qq\Longleftrightarrow\qq 
h_{ij}\to h_{ij}+\pt_{[i}A_{j]}.\eeq
The connection is given by
\beq\label{d5}
\Ga^i_{jk}=\frac 12(g^{-1})_{is}(\pt_j G_{ks}
+\pt_k G_{sj}-\pt_s G_{kj}).\eeq
Using the relation
\beq\label{dmetr}
\pt_i G_{jk}=f_{st}^{~i}\,G_{js}\,G_{tk}=-(GA^iG)_{jk},\eeq
one gets
\beq\label{d6}
\Ga^i_{jk}=-\frac 12f_{st}^{~j}(\Ga^+B^{-1})_{is}\,G_{kt}
-\frac 12f_{st}^{~k}(B^{-1}\Ga^+)_{ti}\,G_{sj}
+\frac 12(g^{-1})_{iu}f_{st}^{~u}\,G_{sj}\,G_{kt}.\eeq
The next step is to simplify the last term in (\ref{d6}). To this end we 
combine Jacobi identity and the definition (\ref{g6}) to prove the identity
\beq\label{d7}
f_{ij}^{~s}\,\Ga^{(\pm)}_{sk}-f_{kj}^{~s}\,\Ga^{(\pm)}_{si}=
2\om_{ik,j}-f_{ik}^{~u}\,\Ga^{(\mp)}_{uj}.\eeq
Starting from relation (\ref{d4}) for the inverse metric we can write
$$(g^{-1})_{iu}\,f_{st}^{~u}=(\Ga^+B^{-1})_{iv}\,f_{st}^{~u}\,\Ga^+_{uv},$$
and use (\ref{d7}) to interchange the indices $s\ \leftrightarrow\ v.$ Several 
simplifications occur then in relation (\ref{d6}) and one is left with the 
simple result
\beq\label{d8}
\Ga^i_{jk}=(f_{is}^{~k}-
\om^t_{~s,u}\,\Ga^+_{it}\,G_{ku})G_{sj}.\eeq
The same procedure, using the second writing of $g^{-1}$ in relation (\ref{d4}), 
gives another interesting form
\beq\label{d9}
\Ga^i_{jk}=(-f_{is}^{~j}+
\om^t_{~s,u}\,\Ga^-_{it}\,G_{uj})G_{ks}.\eeq

\subsection{Torsion}
To get a useful form for the torsion we use relation (\ref{d8}) to compute
$$2T^i_{jk}\,\Ga^+_{jr}\,\Ga^+_{ks}=f_{ir}^{~~k}\Ga^+_{ks}-
\om^{\ga}_{~r,\be}\,\Ga^+_{i\ga}(\Ga^-\,G)_{s\be}-(r\ \leftrightarrow\ s).$$
The identity (\ref{d7}) and the easy relation $\,\Ga^-\,G=2B\,G-{\mb I},$ 
transform the previous relation into
$$T^i_{jk}\,\Ga^-_{rj}\,\Ga^-_{sk}=
-\om^{\ga}_{~r,\be}\,\Ga^+_{i\ga}(B\,G)_{s\be}
-(r\ \leftrightarrow\ s)-\om_{rs,i}.$$
It is natural to multiply both sides by $\,(B\,G)^{-1}_{ur}(B\,G)^{-1}_{ts}.$ 
Observing that $\,g^{-1}=(BG)^{-1}\Ga^-,$ we get
$$T^{ijk}=(\Ga^+B^{-1})_{ks}\,\om^{\ga}_{~s,j}\,\Ga^+_{i\ga}-(j\ \leftrightarrow\ k)
+(\Ga^+B^{-1})_{js}\,(\Ga^+B^{-1})_{kt}\,\om_{rs,i}.$$
This result shows that this tensor is much simpler than $\,T_{ijk}\,$ since it is 
a polynomial in the fields $\,\psi.$ The coefficient of the linear term vanishes 
from Jacobi's identity and we are left with
$$T^{ijk}=\frac 12 f_{ij,k}-(A\cdot\psi)_{i\alf}\,(A\cdot\psi)_{j\be}\,
\om^{\alf\be}_{~~i} +\cdots,$$
where the dots indicate circular permutations of the indices $\,i,\,j,\,k.$ 
We expand the spin connection according to (\ref{g6}) and use the 
identity (\ref{d7}) to end up with
\beq\label{torsgene}
2T^{ijk}=f_{ij,k}
-(A\cdot\psi\,B^{-1})_{it}\,(A\cdot\psi\,B^{-1})_{ju}\,f_{tu,k}
-f_{ij}^{~~s}\,(A\cdot\psi\,B^{-1}A\cdot\psi)_{sk}+\cdots\eeq
Now we can discuss a possibility not yet considered in the literature : the 
vanishing of the torsion in the dual model. The terms which are independent of 
$\,\psi\,$ require $\,f_{[ij,k]}=0,$ a first condition which mixes the structure 
constants and the breaking matrix. Using this relation and the Jacobi identity 
one can check that the last two terms in (\ref{torsgene}) are equal. We conclude 
that the torsion vanishes iff
\beq\label{torsvan}
f_{[ij,k]}=0,\qq\mbox{and}\qq 
f_{\alf s}^{~~(u}\,(B^{-1})_{st}\,f_{t[k}^{~~v)}\,f_{ij]}^{~~\alf}=0,
\quad\forall\ (u,v)\ [ijk].\eeq
Clearly for a simple algebra, the first constraint never holds, but for solvable 
algebras both conditions may be satisfied, as will be seen in section 5 for 
the Bianchi family.

Let us conclude with an example of Lie algebra, for which the torsion 
vanishes for any choice of the breaking matrix. Let its generators be 
$\,\{X_i,\ i=1,\cdots\,\nu\}\,$ and take
$$[X_1,X_i]=X_i,\quad i=2,\cdots,\nu,\qq [X_i,X_j]=0,\quad i\neq j\neq 1.$$

\subsection{Ricci tensor}
The covariant derivatives are defined by 
\beq\label{DC}
D_iv^j=\pt_iv^j+\Ga^j_{is}v^s=\nabla_iv^j+T^j_{is}v^s,
\qq D_iv_j=\pt_iv_j-\Ga_{ij}^sv_s=\nabla_iv_j-T_{ij}^sv_s,\eeq
and the Riemann curvature by
$$[D_k,D_l]v^i={\cal R}^i_{~s,kl}v^s-2T^s_{kl}\,D_sv^i.$$ 
Its explicit form is given by
$${\cal R}^i_{~j,kl}=\pt_k\Ga^i_{lj}-\pt_l\Ga^i_{kj}
+\Ga^i_{ks}\Ga^s_{lj}-\Ga^i_{ls}\Ga^s_{kj}.$$
The Ricci tensor follows from
\beq\label{ric0}
Ric_{ij}={\cal R}^s_{~i,sj}=\pt_s\Ga^s_{ji}-\pt_j\Ga^s_{si}
+\Ga^s_{st}\Ga^t_{ji}-\Ga^s_{jt}\Ga^t_{si}.\eeq
Using
$$\Ga^s_{st}=\ga^s_{st}=\pt_t(\ln\sqrt{\det g}),$$
we get for it a useful form
\beq\label{ric}
Ric_{ij}=\pt_s\Ga^s_{ji}-\Ga^s_{jt}\Ga^t_{si}-D_jD_i(\ln\sqrt{\det g}).\eeq
In order to compute the first two terms in this relation, 
we use (\ref{d8}) for the first two connections and (\ref{d9}) for 
the third one. Apart from trivial cancellations one has to use the identity
\beq\label{id2}
\om^s_{~t,u}\,\Ga^+_{as}+\om^s_{~u,t}\,\Ga^-_{as}=f^{~s}_{at}\,\Ga^-_{su}-
f^{~s}_{ua}\,\Ga^+_{st}\eeq
in order to obtain further strong cancellations of terms, with the final simple 
result
\beq\label{d10}
\pt_s\Ga^s_{ji}-\Ga^s_{jt}\Ga^t_{si}=-G_{is}\,ric_{st}\,G_{tj}
+2f^{s}_{st}\,\om^t_{~u,v}G_{iu}G_{vj}.\eeq
Using (\ref{d9}) and (\ref{DC}), one can check that the last term can be written
$$2f^{s}_{st}\,\om^t_{~u,v}G_{iu}G_{vj}=D_j V_i,\qq\qq 
V_i=-2\,G_{it}\,f^{s}_{st}.$$
Therefore we end up with 
\beq\label{Ric}
Ric_{ij}=-G_{is}\,ric_{st}\,G_{tj}+D_jv_i,\qq\quad
v_i=V_i-\pt_i\ln(\sqrt{\det g}).\eeq
This relation, which displays the relation between the frame geometry of 
the principal model and the geometry of its dual, will play an 
essential role in the next section.

Let us conclude with some remarks :
\brm
\item This result is different, although related to the ones by Tyurin 
\cite{Ty} and Alvarez \cite{Al1}, who expressed the frame geometry of the 
dual model in terms of the frame geometry of the principal model. The 
first reference uses supersymmetry while the second uses purely frames. Our 
approach, using mainly local coordinates computations is valid for any 
breaking matrix $\,B,$ while the previous authors have considered only the 
case $\,B={\mb I}.$ Note also that, in view of the complexity of the 
dualized vielbein it's a long way from the vielbein components of the Ricci 
to our relation (\ref{Ric}).
\item If we consider a simple algebra $\,{\cal G},$ equipped with its 
bi-invariant metric (\ref{g5}). Relation (\ref{E}) shows that the corresponding 
principal model is Einstein and we will prove that the \underline{dual metric 
is quasi-Einstein}. To this aim we insert relation (\ref{E}) into 
(\ref{Ric}), use  $\,f^s_{si}=0\,$ to get for the dual theory
$$Ric_{ij}=\frac{\rho}{4}\,(GBG)_{ij}+D_jv_i.$$
Using relation (\ref{d9}) one can check that
\beq\label{vecteur}
D_j\la_i=\frac 12\,G_{ij}+\frac 12(GBG)_{ij},\qq\qq \la_i=(B^{-1})_{is}\psi^s\eeq
from which we deduce
\beq\label{QE}
Ric_{ij}=-\frac{\rho}{4}\,G_{ij}+D_j\,{\cal V}_i,\qq\quad 
{\cal V}_i=\pt_i\left(-\ln(\sqrt{\det g})+
\frac{\rho}{4}(B^{-1})_{st}\psi^s\psi^t\right),\eeq
which establishes the desired result.

\item One further important point, with respect to string theory, is the dilatonic 
property of the dualized geometry, i.e. wether the vector $V_i$ is a 
gradient or not. For the semi-simple groups the dilatonic property does 
hold since we have $f^s_{st}=0.$
\erm

The failure of this property was first discovered for the dualized Bianchi V 
metric \cite{grv} (see also \cite{egr}). In \cite{gr}, \cite{Ty} it was shown 
to appear when the isometries are not semi-simple and have traceful structure 
constants $\,f^s_{st}\neq 0,$ and its interpretation as an anomaly was 
worked out in \cite{aal1}.

\section{One loop divergences of the dualized models}
We are now in position to discuss the quantum properties of the dualized models 
at the one loop level.

Let us first consider the broken principal models with classical action (\ref{i1}).
Its  one loop counterterm, first computed by Friedan \cite{Fr}, is
\beq\label{r1}
\frac 1{4\pi\eps}\ \int\,d^2x\ ric_{ij}\,\eta^{\mu\nu}J_{\mu}^i\,J_{\nu}^j, 
\qq\quad d=2-\eps,\eeq
where the Ricci components are computed in the vielbein basis.

Renormalizability in the strict field theoretic sense requires that these 
divergences have to be absorbed by (field independent) deformations of the 
coupling constants $\,\hat{\rho}_s\,$ hidden in the matrix $B$ and possibly a 
non-linear field renormalization. The renormalizability 
of the classical theory is ensured by
\beq\label{r2}
ric_{ij}=\hat{\chi}_s(\rho)\frac{\pt}{\pt\hat{\rho}_s}B_{ij}.\eeq

The one loop renormalizability is clear for two extreme choices of metrics :
\brm
\item The bi-invariant metric, for which relation (\ref{E}) shows that the 
principal model is Einstein.
\item The maximally broken metric, for which the matrix $B$ contains 
$\nu(\nu+1)/2$ independent coupling constants $\hat{\rho}_s.$ Since the Ricci 
is also a symmetric matrix, it can always be absorbed by a deformation of 
the coupling constants.
\erm
For partial breakings of the group $G_R,$ relation (\ref{r2}) may fail to 
hold and is indeed a constraint which mixes conditions involving the breaking 
matrix $B$ and the algebra through its structure constants.

In order to compare to the renormalization properties of the dualized theory, 
let us recall that the most general conditions giving one loop 
renormalizability are
\beq\label{renorm1}
\left\{\barr{l}
\dst Ric_{(ij)}=\hat{\chi}_s\,\frac{\pt}{\pt\hat{\rho}_s}\,g_{ij}+
D_{(i}\,u_{j)},\\[3mm]
\dst Ric_{[ij]}=\hat{\chi}_s\,\frac{\pt}{\pt\hat{\rho}_s}\,h_{ij}+
u_s\,T^s_{ij}+\pt_{[i}\,U_{j]},\earr\right.\eeq
where the $\hat{\rho}_s$ are the coupling constants in the principal model we 
started from, appearing now in a non trivial way in the dualized model. The only 
constraint on the functions $\,\hat{\chi}_s\,$ is that they should 
be field independent. 

These relations can be gathered into the single one
\beq\label{renorm2}
Ric_{ij}=\hat{\chi}_s\,\frac{\pt}{\pt\hat{\rho}_s}\,G_{ij}+D_{j}u_{i}+
\pt_{[i}(u+U)_{j]}.\eeq

We are now in position to prove that the one-loop renormalizability of the 
principal model implies the one-loop renormalizability of its dual. For the 
reader's convenience we recall relation (\ref{Ric})
$$Ric_{ij}=-G_{is}\,ric_{st}\,G_{tj}+D_jv_i,\qq\quad
v_i=-2\,G_{it}\,f^{s}_{st}-\pt_i\ln(\sqrt{\det g}),$$
in which we insert (\ref{r2}) to get
$$Ric_{ij}=-\hat{\chi}_l\,G_{is}\frac{\pt}{\pt \hat{\rho}_l}B_{st}\,G_{tj}
+D_jv_i.$$
The first term is reduced using the identity
\beq\label{idder}
\frac{\pt}{\pt\hat{\rho}_l}\,G_{ij}(B,\psi)=
-G_{is}(B,\psi)\,\left(\frac{\pt}{\pt\hat{\rho}_l}B_{st}\right)\,G_{tj}(B,\psi),\eeq
to the final form
\beq\label{resfinal}
Ric_{ij}=\hat{\chi}_s\frac{\pt}{\pt\rho_s}\,G_{ij}+D_jv_i.\eeq
Comparing with relation (\ref{renorm2}) we conclude to the one-loop 
renormalizability of the  dual model. Furthermore the vectors $\,u_i$ and $\,U_i,$ 
defined in relation (\ref{renorm1}), which could be independent, are in fact 
related up to a gauge transformation by $\,U_i=-u_i+\pt_i\tau.$ 

Our next task is to prove that the $\,\be\,$ functions are the same, so  
we need a precise definition of the coupling constants. To do this let us 
switch from the couplings $\,\{\hat{\rho}_i,\ i=1,\ldots,c\}\,$ to new 
couplings $\,(\la,\rho_i)\,$ defined by
\beq\label{couplages}
\hat{\rho}_1=\frac 1{\la},\qq\quad\hat{\rho}_{i+1}=\frac{\rho_i}{\la},
\qq i=1,\ldots c-1.\eeq
We scale similarly the breaking matrix
$$B_{ij}(\hat{\rho})=\frac 1{\la}S_{ij}(\rho),$$
where, for simplicity, the matrix $S$ can be taken linear in 
the couplings $\rho_s.$
Then relation (\ref{r2}) becomes
\beq\label{r3}
\left\{\barr{l}
\dst ric_{ij}(B)=ric_{ij}(S)=
\left(\chi_{\la}+\sum_{s}\chi_s\frac{\pt}{\pt\rho_s}\right)
S_{ij}(\rho),\\[5mm]
\chi_{\la}=\hat{\chi}_1,\qq\chi_i=\hat{\chi}_i-\rho_i\,\hat{\chi}_1,
\qq\qq  i=1,\ldots c-1.\earr\right.\eeq     
The full one loop action is therefore
\beq\label{loop}
\frac 1{\la}\ \frac 12\int\,d^2x
\left[\left(1+\frac{\la\chi_{\la}}{2\pi\eps}\right)S_{ij}(\rho)+
\frac{\la}{2\pi\eps}\sum_{s}\chi_s\frac{\pt}{\pt\rho_s}S_{ij}(\rho)\right]
J_{\mu}^iJ_{\nu}^j,\qq\quad \eps=2-d,\eeq
from which we see that the divergences can be absorbed through coupling 
constant renormalizations :
$$\la_0=\mu^{\eps}\la Z_{\la},\qq Z_{\la}=1-\frac{\la\chi_{\la}}{2\pi\eps},
\qq \rho_i^{(0)}=\rho_iZ_i,\qq\rho_iZ_i=1+\frac{\la\chi_i}{2\pi\eps}.$$
It follows that the corresponding beta functions are
\beq\label{beta}
\be_{\la}=\mu\frac{\pt\la}{\pt\mu}=\la^2\frac{\pt}{\pt\la} Z_{\la}^{(1)}=
-\frac{\la^2}{2\pi}\chi_{\la},\qq
\be_i=\mu\frac{\pt\rho_i}{\pt\mu}=\la\frac{\pt}{\pt\la}(\rho_iZ_i^{(1)})=
\frac{\la}{2\pi}\chi_i.\eeq

For a principal model built with the bi-invariant metric given by (\ref{g5}) 
one has just the single coupling $\la,$ and
\beq\label{betainv}
\be_{\la}=\frac{\la^2}{2\pi}\,\frac{\rho}{4}.\eeq

In order to compute the divergences of the dualized theory in terms of the 
coupling constants defined in (\ref{couplages}) we start from the dual 
classical action
$$G_{ij}(B,\tilde{\psi})\,\pt_+\tilde{\psi}^i\pt_-\tilde{\psi}^j,$$
which we transform according to
$$G(B,\tilde{\psi})=\la\,G(S,\psi),\qq\qq\psi^i=\la\tilde{\psi}^i,\quad
\longrightarrow\quad\frac 1{\la}\,G_{ij}(S,\psi)\,\pt_+\psi^i\,\pt_-\psi^j.$$
The one-loop counterterms follow from the ricci. We start from relation 
(\ref{Ric}) written
$$Ric_{ij}=\la^2\left[-G_{is}(S,\psi)ric_{st}G_{tj}(S,\psi)+D_jv_i\right].$$
Using (\ref{r3}) we write the first term
$$-\chi_{\la} G_{is}(S,\psi)S_{st}G_{tj}(S,\psi)-\sum_u\chi_u G_{is}(S,\psi)
\,\frac{\pt S_{st}}{\pt\rho_u}\,G_{tj}.$$
While the second term is reduced using the identity (\ref{idder}), 
the first term requires more work. One has first to define the vectors
\beq\label{vecpy}
w_i=g_{is}\psi^s,\qq\qq W_i=\psi^sG_{si},\eeq
then check the relation
$$D_jw_i+\pt_{[i}W_{j]}=G_{ij}+
\frac 12\psi^s\left(\pt_jG_{is}+\pt_iG_{sj}\right)-\Ga^t_{ji}g_{ts}\psi^s,$$
which upon use of (\ref{d5}) becomes
$$D_jw_i+\pt_{[i}W_{j]}=G_{ij}+\frac 12\psi^s\pt_sG_{ij}.$$
Eventually relation (\ref{dmetr}) gives
\beq\label{identpy}
D_jw_i+\pt_{[i}W_{j]}=\frac 12G_{ij}+\frac 12(GBG)_{ij}.\eeq

Scaling appropriately this identity, we have
$$\left[\chi_{\la} G_{ij}(S,\psi)+
\sum_u \chi_u\frac{\pt}{\pt\rho_u}\,G_{ij}(S,\psi)+
D_j(v_i-2\chi_{\la}w_i)-2\chi_{\la}\pt_{[i}W_{j]}\right]
\,\pt_+\psi^i\,\pt_-\psi^j.$$
We end up with the renormalized dual theory
\beq\label{final}\left\{\barr{l}
\dst\frac 1{\la}\int\,d^2x\left[\left(1+\frac{\la\chi_{\la}}{2\pi\eps}
+\frac{\la}{2\pi\eps}\sum_s\chi_s\frac{\pt}{\pt\rho_s}\right)G_{ij}
+\frac{\la}{2\pi\eps}\left(D_j{\cal V}_i-2\chi_{\la}\pt_{[i}W_{j]}\right)\right]
\,\pt_+\psi^i\,\pt_-\psi^j,\\[6mm]
\qq{\cal V}_i=v_i-2\chi_{\la}w_i.\earr\right.\eeq

Comparing this relation with (\ref{loop}) we conclude that, up to the non-linear 
field re-definition described by the vector ${\cal V}_i\,$ and the gauge 
transformation  described by $\,W_i,$ the coupling constants renormalization 
are exactly the same as in the principal model we started from. 
We have thus established, at the one-loop level, that the principal $\si-$model 
renormalizability implies the renormalizability, in the strict field theoretic 
sense, of its dual and proved that their $\,\be$ functions do coincide.

\vspace{3mm}

{\bf Remarks :}
\brm
\item What is really new with respect to \cite{Ty} is that, even working with 
renormalizability in the strict field theoretic sense, the (possibly strong) 
breaking of the right isometries $\,G_R$ does not jeopardize the one-loop renormalizability, and even in this extreme sitaution the $\,\be$ functions 
of the principal model and its dual remain the same. This was not obvious since the 
symmetry breaking is a ``hard" breaking, by couplings of power counting dimension two.
\item As already observed in section 2.3, the isometries of $\,G_L$ are lost in 
the dualization process. Hence for the maximal breaking of $G_R,$ no trace seems 
to remain of the original isometries in the dualized theory. These dual theories  constitute a nice example of non-homogeneous metrics with torsion, with no 
isometries to account for their one-loop renormalizability. Our computation,  
which puts forward an experimental fact (the one-loop renormalizability) needs 
some basic theoretical explanation since we know that renormalizability is never accidental but the result of some underlying deeper symmetry.
\item As first observed in \cite{bbfm} for the dualized $\,SU(2)$ model with 
symmetry breaking, there appears in the final form of the divergences (\ref{final}) 
a gauge transformation $\,W_i.$ This term is absent for models built 
on simple Lie groups with their bi-invariant metric $\,B_{ij}.$ Indeed in this 
case we have the identities
$$\psi^sG_{si}=G_{is}\psi^s=(B^{-1})_{is}\psi^s\qq\Longrightarrow\qq 
w_i=W_i=\pt_i\left(\frac 12 (B^{-1})_{st}\psi^s\psi^t\right)\equiv \la_i,$$
which implies $\,\pt_{[i}W_{j]}=0.$  Then the general identity (\ref{identpy}) 
reduces, for this particular case, to relation (\ref{vecteur}).
\item The situation at the two-loop level is still unclear since despite negative 
results in several models \cite{hk1},\cite{bbfm} a more promising and new approach 
to the problem \cite{hk2} seems to yield a positive answer.
\item It is well known that the unbroken principal models are integrable
(for a review see \cite{Ve}). On the contrary the broken ones are not believed 
to be generically integrable, a notable exception being $\,SU(2),$ 
whose integrability was shown in \cite{Ch} for the most general breaking. If 
this belief is confirmed, our 
results show that the one-loop quantum equivalence survives to symmetry 
breaking and therefore the root of this equivalence cannot be integrability.
\erm

\section{Extension to principal models with torsion}
The previous results can be generalized to cover principal models with torsion, 
with action
$$S=\frac 12\,\int\,d^2x\left(B_{ij}\eta^{\mu\nu}+C_{ij}\eps^{\mu\nu}\right)
J_{\mu}^iJ_{\nu}^j,\qq\quad C_{ij}=-C_{ji},$$
where the matrix $\,C$ has {\em constant} components. Taking into account the 
vielbein interpretation of the currents, we  
define the torsion $\,t_{ijk}\,$ as usual by
$$t=\frac 1{3!}\,t_{ijk}e^i\wedge e^j\wedge e^k=\frac 12\, dC,\qq\quad 
C=\frac 12\,C_{ij}e^i\wedge e^j,$$
which gives
$$t_{ijk}=-\frac 12\left(f_{ij}^{~~s}C_{sk}+f_{jk}^{~~s}C_{si}
+f_{ki}^{~~s}C_{sj}\right).$$

One should first observe that the parallelizing torsion (\cite{Wi}) is not 
of this kind, and second that we have to exclude the case where
\beq\label{label}
C_{ij}=f_{ij}^{~~s}\,\ga_s.\eeq
Indeed, if this relation holds the Bianchi identity (\ref{b}) gives
$$C_{ij}\eps^{\mu\nu}J_{\mu}^iJ_{\nu}^j=
\ga_s\,f_{ij}^{~~s}\,\eps^{\mu\nu}\,J_{\mu}^i\,J_{\nu}^j=
-2\ga_s\,\eps^{\mu\nu}\,\pt_{\mu}J_{\nu}^s,$$
which is a total divergence. Correspondingly the torsion vanishes as a 
consequence of the Jacobi identity.

Even if (\ref{label}) is valid for a semi-simple algebra $\,{\cal G},$ it is 
not valid for {\em any} algebra. To see this let us suppose that the center of
$\,{\cal G},$ is non-trivial, i. e. there is some generator $\,X_{\alf}$ 
which commutes with all the other generators. It follows that 
$\,f_{\alf i}^{~s}\,\ga_s\equiv 0\,$ for all values of $\,i,$ while $\,C_{\alf i}$ 
can be non-vanishing.

Let us describe briefly how our analysis can be generalized.

The spin connection $\,\Om^i_{~j}\,$ now verifies
$$de^i+\Om^i_{~j}\wedge e^j=B^{ij}t_j,\qq t_i=t_{ist}\,e^s\wedge e^t.$$
Let us define
$$\Om^{(\pm)}_{~~~ij,k}=\om_{ij,k}\pm t_{ijk},\qq \Om^{(\pm)~i}_{~~~~~j,k}=
(B^{-1})_{is}\Om^{\pm}_{~~sj,k},$$
then the spin connection one-forms are $\,\Om^i_{~j}=\Om^{(-)~i}_{~~~~~j,s}\,e^s.$

The Ricci tensor has now for components
$$ric_{ij}=\Om^{(+)~s}_{~~~~~t,s}\,\Om^{(-)~t}_{~~~~~i,j}
-\Om^{(-)~s}_{~~~~~i,t}\,\Om^{(+)~t}_{~~~~~j,s},\qq \Om^{(+)~s}_{~~~~~t,s}=
\Om^{(-)~s}_{~~~~~t,s}$$
and is no longer symmetric.

Introducing the notations $\,\Ga^{\pm}=B\pm(C+A\cdot\psi),$ we have for the 
dualized metric  $\,G=(\Ga^+)^{-1}.$ The connection in the dual theory 
becomes
$$\Ga^i_{jk}=(f_{is}^{~k}-
\Om^{(+)~t}_{~~~~~s,u}\,\Ga^+_{it}\,G_{ku})G_{sj}=(-f_{is}^{~j}+
\Om^{(-)~t}_{~~~~~s,u}\,\Ga^-_{it}\,G_{uj})G_{ks},$$
from which, after tedious computations, one gets for the Ricci tensor
\beq\label{RicT}
Ric_{ij}=-G_{is}\,ric_{st}\,G_{tj}+D_jv_i,\qq\quad
v_i=-2\,G_{it}\,f^{s}_{st}-\pt_i\ln(\sqrt{\det g}),\eeq
which is strikingly similar to (\ref{Ric}).

Let us denote by $\,\rho^B_s\,$ the couplings present in the matrix $\,B,$ by 
$\,\rho^C_s\,$ the couplings present in the matrix $\,C,$ and $\,\rho_s\,$ 
the couplings present in both matrices. The 
renormalizability of the principal model with torsion is ensured by
\beq\label{ren}
\left\{\barr{l}
\dst ric_{(ij)}=\left(\chi_s\,\frac{\pt}{\pt \rho^B_s}
+\eta_s\,\frac{\pt}{\pt \rho_s}\right)B_{ij},\\[4mm]
\dst ric_{[ij]}=\left(\eta_s\,\frac{\pt}{\pt \rho_s}
+\xi_s\,\frac{\pt}{\pt \rho^C_s}\right)C_{ij}.\earr\right.\eeq
Inserting relation (\ref{ren}) into (\ref{RicT}) one ends up with
\beq\label{renT}
Ric_{ij}=\left(\chi_s\,\frac{\pt}{\pt \rho^B_s}+\eta_s\,\frac{\pt}{\pt \rho_s}
+\xi_s\,\frac{\pt}{\pt \rho^C_s}\right)G_{ij}+D_jv_i.\eeq
It follows, by the same arguments as in section 5, that the dual model is also 
renormalizable  and has the same $\,\be\,$ functions as the initial principal 
model with torsion.

\section{Dualized Bianchi metrics}
Particular dualized models in the Bianchi family have been studied with 
emphasis either put on the renormalizability properties of the dualized 
models with symmetry breaking \cite{bbfm}, \cite{hk1} or on the dilaton 
anomaly \cite{grv}, \cite{egr}. The aim of this section is to give some 
detailed analysis of both aspects for the full family.

All the Lie algebras with 3 generators were classified by Bianchi (1897). In a 
modern presentation \cite{ewb}, \cite{emc} these algebras are described in 
terms of the parameter $a$ and the vector $\vec{n}=(n_1,n_2,n_3)$ according to
$$[X_1,X_2]=aX_2+n_3X_3,\qq[X_2,X_3]=n_1X_1,\qq [X_3,X_1]=n_2X_2-aX_3,
\quad f^s_{st}=-2a\de_{t1}.$$
The Jacobi identity requires $\,a\cdot n_1=0.$

The algebras of interest appear in the following table

\vspace{5mm}
\centerline{
\begin{tabular}{|l|c|c|c|c|l|c|c|c|c|}
\hline
Class A : $\,a=0\ \ $ & & & & & Class B : $\,n_1=0,\ \,a>0\ \ $ & & & &   \\
\hline\hline
type    &  & $n_1$ & $n_2$ & $n_3$ & \quad type  & & $a$ & $n_2$ & $n_3$    \\
\hline
I  &  & 0 & 0 & 0            & \quad V       & & 1         & 0   &  0 \\
II &  & 1 & 0 & 0            & \quad IV      & & 1         & 0   &  1 \\
VI$_0$  &  & 0   & 1   & -1  & \quad III     & & 1         & 1   & -1 \\   
VII$_0$ &  & 0     & 1   & 1 & \quad VI$_a$  & & $a\neq 1$ & 1   & -1 \\   
VIII $\quad su(1,1)$ &  & 1   & 1     &-1    &\quad VII$_a$ & &   & 1     &1 \\  
IX $\quad\ \ su(2)$  &  & 1     & 1     & 1  & & & & & \\   
\hline  
\end{tabular}}
\vspace{5mm}
\noindent The adjoint representation is given by
\beq\label{adj}
T_1=\left(\barr{ccc} 0 & 0 & 0\\
0 & -a & -n_3\\ 0 & n_2 & -a\earr\right),\qq T_2=
\left(\barr{ccc} 0 & a & n_3\\ 0 & 0 & 0\\ -n_1 & 0 & 0\earr\right),
\qq T_3=
\left(\barr{ccc} 0 & -n_2 & a\\ n_1 & 0 & 0\\ 0 & 0 & 0\earr\right).\eeq

The Killing metric $\,g_{ij}=\,{\rm Tr}\,(T_iT_j)\,$ is diagonal with 
$$g_{11}=2(a^2-n_2n_3),\qq g_{22}=-2n_3n_1,\qq g_{33}=-2n_1n_2.$$
It follows that  B VIII and B IX are semi-simple (in fact, simple). Among the 
remaining non semi-simple algebras, only those in class B have traceful 
structure constants.

To simplify matters, still keeping the main peculiarities of symmetry breaking, 
we take the diagonal metric $\,B_{ij}=r_i\,\de_{ij}.$ The dual metric tensor 
is then
\beq\label{dualB}
G=\frac 1{\Delta_+} \left(\barr{ccc}
r_2r_3+x^2 & r_3z-xy & -r_2y-zx\\[3mm]
-r_3z-xy & r_3r_1+y^2 & -r_1x+yz\\[3mm]
r_2y-zx & r_1x+yz & r_1r_2+z^2\earr\right),\qq\quad\left\{
\barr{l} x=n_1\psi^1,\\[3mm] y=n_2\psi^2-a\psi^3,\\[3mm]
 z=a\psi^2+n_3\psi^3.\earr\right.\eeq
with
$$\Delta_{\pm}=r_1r_2r_3+r_1x^2\pm (r_2y^2+r_3z^2).$$ 

From (\ref{torsion}) we get the torsion
\beq\label{tor}
\left\{\barr{l}
\dst T_{ijk}=t\,\eps_{ijk},\qq\qq\quad t=\frac{N}{\Delta_+^2},\\[5mm]
N=\nu\Delta_- +2r_2r_3(n_3y^2+n_2z^2-n_1r_1^2),\qq\qq 
\nu=r_1n_1+r_2n_2+r_3n_3.
\earr\right.\eeq
This result shows that for Bianchi V the dualized metric 
is torsion free !

Relation (\ref{ricci}) gives for the non-vanishing vielbein components of the 
initial Ricci tensor 
\beq\label{ricciB}
\left\{\barr{ll}
\dst ric_{11}=-2a^2+\frac{n_1^2r_1^2-(n_2r_2-n_3r_3)^2}{2r_2r_3}, &  \\[4mm]
\dst ric_{22}=-2a^2\,\frac{r_2}{r_1}+\frac{n_2^2r_2^2-(n_3r_3-n_1r_1)^2}{2r_3r_1}, & 
\dst\qq ric_{23}=ric_{32}=a\,\frac{(n_2r_2-n_3r_3)}{r_1}, \\[4mm]
\dst ric_{33}=-2a^2\,\frac{r_3}{r_1}+\frac{n_3^2r_3^2-(n_1r_1-n_2r_2)^2}{2r_1r_2}.
\earr\right.\eeq

\subsection{Class A dual models and their $\be$ functions}
We see at a glance from the Ricci that the class A principal models, with 3 
independent diagonal couplings are renormalizable at one-loop. 
From the the previous section this ensures the renormalizability of the 
dualized model, with the same $\,\be\,$ functions. 

For Bianchi I and II we define
$$r_1=\frac 1{\la},\qq r_2=\frac{g}{\la},\qq r_3=\frac{g'}{\la}.$$
Then one gets for the $\,\be$ functions
$$\left\{\barr{l}
\dst\mbox{Bianchi I}\ :\quad \be_{\la}=\be_g=\be_{g'}=0\\[4mm]
\dst\mbox{Bianchi II}\ :\quad \be_{\la}=-\frac{\la^2}{4\pi}\frac 1{gg'},\quad 
\be_{g}=-\frac{\la}{8\pi}\,\frac 1{g'},\quad \be_{g'}=
-\frac{\la}{8\pi}\,\frac 1{g}.\earr\right.$$
The result for Bianchi I is obvious, since its metric is flat.

For Bianchi IX (with $\,\si=+1\,$) and  Bianchi VIII (with  $\,\si=-1\,$), we 
parametrize the couplings according to \footnote{In the $\,g=g'=0$ limit we
 recover the bi-invariant metrics.}
$$r_1=\frac{\si}{\la},\qq r_2=\frac{\si(1+g)}{\la},\qq r_3=\frac{1+g'}{\la}.$$
With three independent couplings, the 
$\,SU(2)_R\,$ isometries are fully broken. If one takes $\,g=g'\,$ the 
corresponding model has a residual $\,U(1)_R\,$ isometry and has been studied 
in \cite{bbfm},where the quantum equivalence was proved at the one-loop order.

Using (\ref{ricciB}) and (\ref{beta}) it is a simple matter to compute
\beq\label{betaBsimple}
\left\{\barr{l}
\dst\be_{\la}=-\frac{\la^2}{4\pi}\frac{(1+g-g')(1-g+g')}{(1+g)(1+g')},\\[4mm]
\dst\be_{g}=\frac{\la}{2\pi}\frac{g(1+g-g')}{(1+g')},\qq
\be_{g'}=\frac{\la}{2\pi}\frac{g'(1-g+g')}{(1+g)}.\earr\right.
\eeq
For $\,g=0\,$ and $\,\si=1$ these results agree with \cite{bbfm}.

For the remaining models we parametrize the couplings according to
$$r_1=\si\frac 1{\la},\qq r_2=\si\frac{1+g}{\la},\qq r_3=\frac{1+g'}{\la},$$
where $\,\si=+1\,$ (resp. $\,\si=-1\,$) correspond to 
Bianchi VII$_0$ (resp. Bianchi VI$_0$). We get for the $\be$ functions
\beq\label{betaBVIa}
\be_{\la}=\frac{\la^2}{4\pi}\frac{(g-g')^2}{(1+g)(1+g')},\qq 
\be_g=\frac{\la}{2\pi}\frac{(1+g)}{(1+g')}(g-g'),
\qq\be_{g'}=-\frac{\la}{2\pi}\frac{(1+g')}{(1+g)}(g-g').\eeq
Let us observe that for $\,g'=g\,$ the metric of the principal model is flat, 
which explains the vanishing of all the $\,\be$ functions.

\subsection{Class B dual models and their $\be$ functions}
Let us begin with Bianchi V, which has diagonal ricci, and is therefore 
renormalizable with three independent couplings 
$$r_1=\frac 1{\la},\qq r_2=\frac{1+g}{\la},\qq r_3=\frac{1+g'}{\la}.$$
One gets
\beq\label{betaBV}
\be_{\la}=\frac{\la^2}{\pi},\qq \be_g=\be_{g'}=0.\eeq

For the remaining models in this class the ricci is not diagonal,  
therefore we conclude to the non-renormalizability of the remaining models 
with three independent couplings.

However, if we restict ourselves to two couplings, tuned in such a way to have 
$\,ric_{23}=0,$ most of the class B models become renormalizable:
$$\left\{\barr{lccc}
\mbox{Bianchi III} \qq & \dst r_1=\frac 1{\la}, & \dst r_2=\frac{1+g}{\la}, & 
\dst r_3=-\frac{1+g}{\la}.\\[4mm]
\mbox{Bianchi VI}_a(\si=-1),\ \mbox{VII}_a(\si=+1) \qq & \dst r_1=\frac 1{\la}, 
& \dst r_2=\frac{1+g}{\la}, & \dst r_3=\si\frac{1+g}{\la}.\earr\right.$$
Their beta functions are 
\beq\label{betaBVI}
\be_{\la}=\frac{\la^2}{\pi}\,a^2,\qq\be_g=0.\eeq

For Bianchi IV no choice of diagonal breaking matrix leads to renormalizability.

\subsection{Class B dual models and dilaton anomaly}
Let us first get a convenient characterization of the absence 
of the dilaton anomaly. Using relation (\ref{dmetr}) one has the equivalence
$$V_i=-2G_{it}f^s_{st}=\pt_i\Phi\quad\Longleftrightarrow\quad 
\pt_i\,V_j-\pt_j\,V_i=0
\quad\Longleftrightarrow\quad 
G_{su}f^v_{vu}(f^i_{st}G_{jt}-f^j_{st}G_{it})=0.$$
Upon multiplication by $\,\Ga_{ai}\,\Ga_{bj}\,$ and use of (\ref{id1}), 
(\ref{id2}) one gets
$$G_{su}f^v_{vu}(f^t_{sb}\Ga_{at}-f^t_{sa}\Ga_{bt})=0,\qq\longrightarrow\qq
\om_{ab,s}G_{st}f^u_{ut}=0.$$
It follows that the equivalence becomes
\beq\label{test}
V_i=-2G_{it}f^s_{st}=\pt_i\Phi\quad\Longleftrightarrow\quad
\om_{ab,s}V_s=0,\qq\forall a,b.\eeq
Despite the convenient form of the final relation (\ref{test}), it is fairly 
difficult to discuss in general. Let us simply observe that 
the matrices $\om_a,$ with matrix elements defined by $(\om_a)_{bs}=\om_{ab,s}$ 
are singular. So the analysis of (\ref{test}) depends strongly on the size 
of the kernel of the $\om_a$ and therefore of the algebra and of the breaking 
matrix considered.

To discuss this point for the class B of the Bianchi family, we will consider 
the most general breaking matrix $\,B\,$ and we denote its off-diagonal 
terms by
$$B_{12}=s_3,\quad B_{23}=s_1,\quad B_{31}=s_2,\qq 
\det B=r_1r_2r_3-r_1s_1^2-r_2s_2^2-r_3s_3^2+2s_1s_2s_3\neq 0.$$
Let us notice that this last condition forbids the simultaneous vanishing of 
$s_1,\,r_2\,$ and $\,r_3.$

The matrices $\om_a$ are given generally by
$$\left(\om_i\right)_{jk}=\om_{ij,k}=-\frac{\nu}{2}\,\eps_{ijk}+
\sum_sn_sB_{sk}\eps_{sij}+a_iB_{jk}-a_jB_{ik},\qq a_i=a\de_{i1},
\qq\nu=\sum_s n_sB_{ss}.$$
For class B we have $\ \nu=n_2r_2+n_3r_3.\ $
Taking into account the relations
$$G_{11}=\frac{(r_2r_3-s_1^2)}{\det\,\Ga},\quad
G_{21}=-\frac{(r_3s_3-s_1s_2+s_1y+r_3z)}{\det \Ga},\quad
G_{31}=\frac{(s_3s_1-r_2s_2+r_2y+s_1z)}{\det \Ga},$$
$$\det\Ga=\det B+r_2y^2+r_3z^2+2s_1yz,$$
it is a purely algebraic matter, using (\ref{test}), to prove that the dilaton 
anomaly is absent iff
\beq\label{testB}
\nu\equiv n_2r_2+n_3r_3=0\qq\mbox{and}\qq \mu\equiv s^2_1-r_2r_3=0.\eeq
These constraints show that Bianchi VII$_a$ is always anomalous, but also that 
an appropriate choice of the couplings can get rid of the anomaly in the 
other models!

One can summarize the constraints (\ref{testB}) for the class B models and 
their possibly non-vanishing ricci component :

\vspace{5mm}
\centerline{
\begin{tabular}{|l|c|c|c|}
\hline
model & constraint\quad$(\eps=\pm 1)$ & $ric_{11}$ & $\det B\neq 0$ \\
\hline\hline
Bianchi III & $r_3=r_2,\ s_1=\eps r_2$ & $2(\eps-1)$ & $r_2(s_2-\eps s_3)\neq 0$\\
Bianchi IV & $r_3=0, s_1=0$ & 0 & $r_2\cdot s_2\neq 0$\\
Bianchi V\qq & $s_1=\eps\sqrt{r_2r_3}$,  & 0 & 
$\ \ \sqrt{|r_2|}s_2-\eps\sqrt{|r_3|}s_3\neq 0\ \ $ \\
Bianchi VI$_a$ ($a\neq 1$)$\quad$ & $\ \ r_3=r_2,\ s_1=\eps r_2\ \ $ & $\ \ 2(a\eps-1)\ \ $ & 
$r_2(s_2-\eps s_3)\neq 0$\\
\hline
Bianchi VII$_a$ & impossible & & \\
\hline
\end{tabular}}

\vspace{5mm}

\noindent It follows that the models  B IV, B V and B III with $\,\eps=+1$ 
are flat.

We want to show that the restrictions (\ref{testB}) are equivalent to the 
vanishing of the torsion. To see this  we use the constraints 
(\ref{torsvan}), which give, when specialized to class B :
$$3\,f_{[ij,k]}=\nu,\qq\qq
3\,(A\cdot\psi\,B^{-1}A\cdot\psi)_{s[k}\,f_{ij]}^{~~s}=
\mu\ \frac{(a^2+n_2n_3)(n_2(\psi^2)^2+n_3(\psi^3)^2)}{\det B}.$$

In this case it is interesting to compare the vectors $\,V_i=-2G_{it}f^s_{st}\,$ 
and $\,g_i=D_i\ln(\sqrt{\det g}).$ One can check that the difference 
$\,V_i-2g_i\,$ is then covariantly constant, giving for final geometry 
$$Ric_{ij}=-G_{is}\,ric_{st}\,G_{tj}+D_jD_i\ln(\sqrt{\det g}).$$

\section{Dualized Bianchi V model at two loops}
As observed in the previous sections, dualized models may be {\em torsionless} : 
it is therefore important to ascertain which models lead to this 
phenomenon. To this end we use the constraints (\ref{torsvan}). 
Algebraic computations lead to the following conclusions:
\brm
\item For the class A models, no choice of the non-singular matrix $\,B$ leads 
to vanishing torsion.
\item For the class B models, except Bianchi V, the necessary and sufficient 
conditions for vanishing torsion are given by the relations (\ref{testB}).
\item Among all the class B models only Bianchi V has a vanishing torsion for 
an arbitrary breaking matrix $\,B.$ In this case the torsion potential is an 
exact 2-form with 
$$\left\{\barr{l}
\dst H=\frac 12\, dA,\qq\qq\nu_2=\frac{s_3s_1-r_2s_2}{s_1^2-r_2r_3},\qq 
\nu_3=\frac{s_1s_2-r_3s_3}{s_1^2-r_2r_3},\\[6mm]
\dst A=\ga\left(d\psi^1-\nu_3\,d\psi^2-\nu_2\,d\psi^3\right),
\qq\ga=\ln(\sqrt{\det g}).\earr\right.$$
\erm
The case where $\,s_1^2=r_2r_3\neq 0\,$ is special, with
$$A=\ga d\psi^1+\frac 1{r_2}\left(s_3\ln|x^2-\alf^2|
-\ln\left|\frac{x+\alf}{x-\alf}\right|\cdot\psi^2\right)d\psi^2,\quad
x=r_2\psi^3-s_1\psi^2,\quad\alf=s_1s_3-r_2s_2.$$
It follows that the dual model, at least perturbatively, can be analyzed as if 
it had no WZW coupling ! This situation is fairly original : the principal 
Bianchi V model, which is homogeneous and torsionless, is mapped by T-duality 
to an inhomogeneous but still torsionless $\,\si$-model. It is therefore attractive 
to check the two-loop equivalence of the models using the firmly 
established counterterms given by Friedan \cite{Fr}. Let us consider the 
simplest Bianchi V dual model, with $\,B_{ij}=r\de_{ij}.$ Its dualized metric, 
taken from (\ref{dualB}), reads :
\beq\label{gBV}
g=\frac r{\De}(d\psi^1)^2+
\frac 1{\De}\left[r^2(d\psi^2)^2+r^2(d\psi^3)^2+
(\psi^3d\psi^2-\psi^2d\psi^3)^2\right],\qq\De=(\psi^2)^2+(\psi^3)^2+r^2.\eeq
Following \cite{egr} we take for new coordinates
\beq\label{gBVmod}
\psi^1=z,\ \ \psi^2+i\psi^3=\rho\,e^{i\phi},\qq\Longrightarrow\qq 
g=\frac{r}{\rho^2+r^2}(dz^2+d\rho^2)+\frac{\rho^2}{r}\,(d\phi)^2,\eeq
which bring the metric to a simple diagonal form, with the obvious vielbein
\beq\label{dreibein}
\dst g=\sum_{a=1}^3e_a^2,\qq e_1=\frac{\sqrt{r}}{\sqrt{\rho^2+r^2}}\,dz,\quad
e_2=\frac{\sqrt{r}}{\sqrt{\rho^2+r^2}}\,d\rho,\quad 
e_3=\frac{\rho}{\sqrt{r}}\,d\phi.\eeq
One can prove that this metric has two isometries, described by 
the vector fields $\dst\,\frac{\pt}{\pt z}\,$ and $\dst\,\frac{\pt}{\pt\phi}.$

The geometrical quantities of interest are
\beq\label{geo}\left\{\barr{l}
\dst\om_{23}=-\frac 1{\sqrt{r}}\frac{\sqrt{\rho^2+r^2}}{\rho}\,e_3,
\qq\qq\om_{12}=-\frac 1{\sqrt{r}}\frac{\rho}{\sqrt{\rho^2+r^2}}\, e_1,\\[5mm]
\dst R_{23}=-\frac 1r\,e_2\wedge e_3,\qq\quad R_{31}=\frac 1r\,e_3\wedge e_1,
\qq\quad R_{12}=-\frac{\si}{r}\,e_1\wedge e_2,\qq 
\si=\frac{\rho^2-r^2}{\rho^2+r^2},\\[5mm]
\dst Ric_{11}=\frac{1-\si}{r},\qq Ric_{22}=-\frac{1+\si}{r},\qq Ric_{33}=0,\\[5mm]
\dst R=Ric_{ss}=-2\,\frac{\si}{r}.\earr\right.\eeq
The one-loop renormalizability relations 
$$Ric_{ij}=\chi^{(1)}\frac{\pt}{\pt r}\,g_{ij}+\nabla_{(i}v_{j)},$$
become, using vielbein components
\beq\label{1boucle}\left\{\barr{l}
\dst Ric_{ab}=\chi^{(1)}\left((e^{-1})_b^j\,\frac{\pt}{\pt r}\,e_{aj}+
(e^{-1})_a^j\,\frac{\pt}{\pt r}\,e_{bj}\right)+{\cal D}_{(a}v_{b)},\\[5mm]
\dst {\cal D}_av_b=\hat{\pt}_a\, v_b+\om_{bs,a}\,v_s,
\qq\quad \hat{\pt}_a=(e^{-1})_a^j\,\pt_i.\earr\right.\eeq
Relation (\ref{1boucle}) works with
\beq\label{renorm}
\chi^{(1)}=-2, \qq\qq 
v\equiv v_a\,e_a=-\frac 2{\sqrt{r}}\frac{\rho}{\sqrt{\rho^2+r^2}}\,d\rho.\eeq
Let us remark that while $\,\chi^{(1)}\,$ is uniquely defined, the vector 
$\,v_i$ is not unique and we took its simplest form. 
As it should, the renormalization of the coupling constant 
$\,r$ is the same as in the principal model as can be seen from 
relation (\ref{ricciB}).

The two-loops counterterms, first computed by Friedan \cite{Fr}, are
$$\frac 1{16\pi^2\eps}\,\int\,d^2x\ R_{is,tu}\,R_{js,tu}\,
\eta^{\mu \nu}\,J_{\mu}^i\,J_{\nu}^j,$$
where the $\,R_{is,tu}\,$ are the vielbein components of the Riemann tensor.

For three dimensional geometries, this counterterm is most easily obtained 
from the identity 
\beq\label{id3}
(RR)_{ab}\equiv\frac 12R_{as,tu}\,R_{bs,tu}=R\,Ric_{ab}-(Ric^2)_{ab}+
\left({\rm Tr}\,(Ric^2)-\frac{R^2}{2}\right)\de_{ab},\eeq
which gives
$$(RR)_{11}=(RR)_{22}=\frac 1{r^2}(1+\si^2),\qq\qq (RR)_{33}=\frac 2{r^2}.$$
In order to prove renormalizability we have to solve for $\,\chi^{(2)}$ and 
$\,w_a$ such that
\beq\label{renorm2b}
(RR)_{ab}=\chi^{(2)}\left((e^{-1})_b^j\,\frac{\pt}{\pt r}\,e_{aj}+
(e^{-1})_a^j\,\frac{\pt}{\pt r}\,e_{bj}\right)+{\cal D}_{(a}w_{b)}.\eeq
Explicitly, these equations give the differential system
\beq\label{syst1}\left\{\barr{ll}
\dst\frac 1{r^2}(1+\si^2)-\chi^{(2)}\frac{\si}{r}=\hat{\pt}_1\,w_1+\om_{12,1}\,w_2,
\qq\quad & 0=\hat{\pt}_1\,w_2+\hat{\pt}_2\,w_1-\om_{12,1}\,w_1,\\[3mm]
\dst\frac 1{r^2}(1+\si^2)-\chi^{(2)}\frac{\si}{r}=\hat{\pt}_2\,w_2,
\qq\quad & 0=\hat{\pt}_3\,w_1+\hat{\pt}_1\,w_3,\\[3mm]
\dst\hspace{15mm}\frac 2{r^2}+\chi^{(2)}\frac 1r=\hat{\pt}_3\,w_3-\om_{23,3}\,w_2,
\qq\quad & 0=\hat{\pt}_3\,w_2+\hat{\pt}_2\,w_3+\om_{23,3}\,w_3.\earr\right.\eeq
Integrating some relations with respect to the variable $\,\rho$ we obtain
\beq\label{syst2}
\left\{\barr{l}
\dst w_1=-w_1^{(0)}(\rho)\,\pt_z\,W_2(z,\phi)
+\frac{W_1(z,\phi)}{\sqrt{\rho^2+r^2}},\qq
\frac{d}{d\rho}\left(\sqrt{\rho^2+r^2}\,w_1^{(0)}(\rho)\right)=\sqrt{\rho^2+r^2},
\\[4mm]
\dst w_2=w_2^{(0)}(\rho)+W_2(z,\phi),\qq\qq\frac{d}{d\rho}w_2^{(0)}(\rho)=
\frac{\sqrt{r}}{\sqrt{\rho^2+r^2}}
\left(\frac{1+\si^2}{r^2}-\chi^{(2)}\,\frac{\si}{r}\right),\\[4mm]
\dst w_3=-\frac{\sqrt{\rho^2+r^2}}{r}\,\pt_{\phi}\,\,W_2(z,\phi)+
\rho\,W_3(z,\phi).\earr\right.\eeq
Inserting these relations into the last left relation of (\ref{syst1}) one has
\beq\label{fin}
\frac 2{r^2}+\frac{\chi^{(2)}}{r}-\sqrt{r}\pt_{\phi}\,W_3=M,\qq W_2-\pt^2_{\phi}\,W_2=N,\qq 
w_2^{(0)}(\rho)+N=\sqrt{r}\frac{\rho}{\sqrt{\rho^2+r^2}}\,M,\eeq
where $\,M$ and $\,N$ are coordinate independent. Differentiating this last 
relation with respect to $\,\rho$ yields a constraint which does not hold, 
irrespectively of the values taken for $\,M$ and $\,\chi^{(2)}.$

The failure of relations (\ref{renorm2b}) means that the two-loops quantum 
extension chosen for the dual model does not lift the classical equivalence 
to the quantum level. 

In fact we should consider \footnote{We thank G. Bonneau 
for suggesting to us this idea.} the whole family of metrics 
$$g_{ij}\qq\longrightarrow\qq g_{ij}+\ga_{ij},$$
where $\,\ga_{ij}\,$ is a one-loop deformation of the classical metric 
$\,g_{ij}\,$ which describes different possible quantum extensions of the 
same classical dual model. For this modified theory we get an extra 
contribution at the two-loops level which is
$$\frac 1{4\pi\eps}\,\int\,d^2x\,\left[Ric_{ij}(g+\ga)-Ric_{ij}(g)\right]
\,\eta^{\mu \nu}\,J_{\mu}^i\,J_{\nu}^j.$$

Let us examine whether the two-loops renormalizability can be implemented 
or not. As is well known, one has
$$Ric_{ij}(g+\ga)-Ric_{ij}(g)=-\frac 12\,\Delta_L\,\ga_{ij}+
\nabla_{(i}\alf_{j)},\qq\alf_i=\nabla^s\,\ga_{si}-\frac 12\nabla_i\,\ga^s_{~s},$$
where $\,\Delta_L\,$ is Lichnerowicz's laplacian
$$\Delta_L\,\ga_{ij}=\nabla^s\,\nabla_s\,\ga_{ij}+2R_{is,jt}\,\ga^{st}
-Ric_{is}\,\ga^s_{~j}-Ric_{js}\,\ga^s_{~i}.$$
The connection $\,\nabla,$ the Riemann tensor and the raising or lowering 
of indices are related to the unperturbed metric $\,g.$

Up to a scaling of $\,\ga,$ the two-loops renormalizability constraints become 
\beq\label{2loop}
(RR)_{ij}=\Delta_L\,\ga_{ij}+\chi^{(2)}\frac{\pt}{\pt r}\,g_{ij}+
\nabla_{(i}w_{j)}.\eeq
We will exhibit a solution of these equations for any choice of $\,\chi^{(2)}.$ 
For this we consider the vielbein components of the deformation
$$\ga=\frac 1r\left(\ga_1\,e_1^2+\ga_2\,e_2^2+\ga_3\,e_3^2\right),$$
and we use the notations
$$\chi^{(2)}=\frac 2r(1-\chi), \qq x=\frac{\rho^2}{\rho^2+r^2}\,\in\,[0,1[,\qq
\Phi(x)=\frac{3+\chi}{2(1-x)^2}\,\int_0^x\,\frac{\ln(1-u)}{u}\,du.$$
One should notice that for the principal Bianchi V model at two-loops 
we have $\,\chi=0.$

Let us define
$$\left\{\barr{l}
\dst \ga_1(x)=-\frac{1+x}{1-x}\ga_2(0)-\frac{4(4+3\chi)x-(5\chi-1)x^2}{8(1-x)^2}
-\frac{3+\chi-x}{2(1-x)}\ln(1-x)+\,\Phi(x),\\[4mm]
\dst \ga_2(x)=\ga_2(0)+\frac{4(11+4\chi)x+(11+5\chi)x^2-4x^3}{8(1-x)^2}\\[4mm]
\dst \hspace{7cm}+\frac{8+3\chi+x}{2(1-x)}\,\ln(1-x)-(1+2x)\,\Phi(x),\\[4mm]
\ga_3(x)=-\ga_1(x).\earr\right.$$
The vector vielbein components are
$$w_1=w_3=0,\qq r^{3/2}\sqrt{x}\,w_2=4(1-\chi)x-2x^2
-(2+\chi)\ln(1-x)-2x(1-x)\frac{d}{dx}\,\ga_2(x).$$
The reader can check that the deformation and the vector given above are indeed 
solution of (\ref{2loop}) for {\em any value} of $\,\chi.$ 

Two main points need to be checked. The first one is the analyticity of 
the $\,\ga_i(x)\,$ in a neighbourhood of $\,x=0.$ This follows from the 
analyticity of $\,\Phi(x)\,$ and is explicit on the other terms. The 
second point is that we are using polar coordinates ; in order to secure 
an analytic dependence with respect to the cartesian coordinates  $\,\psi^1,\,\psi^2\,\approx 0\,$ we have 
imposed $\,\ga_3(0)=\ga_2(0).$ The free parameters in this solution 
are $\,\ga_2(0)\,$ and $\,\chi.$

As a side remark, let us observe that the deformation obtained above, cannot 
be written in the form
$$\ga_{ij}=A\,\frac{\pt}{\pt r}\,g_{ij}+D_{(i}{\cal W}_{j)},$$
which means that it cannot be interpreted as a {\em finite} renormalization 
of the initial metric $\,g_{ij}.$

So we can conclude that it is always possible to have a quantum extension 
of the dualized Bianchi V which does preserve the two-loops renormalizability. 
Unfortunately nothing, in this process, enforces $\,\chi\,$ to have the 
same value as in the principal model we started from. This shows that further 
constraints are needed to define uniquely the two-loops quantum dual theory.

\vspace{3mm}

\noindent{\bf Acknowledgments :} We are indebted to O. Alvarez, G. Bonneau,  
F. Delduc and E. Ivanov for enlightening discussions.

\end{document}